
\input epsf
\input phyzzx
\overfullrule=0pt
\hsize=6.5truein
\vsize=9.0truein
\voffset=-0.1truein
\hoffset=-0.1truein

\def\scripm{{\cal I}_{\pm}}

\rightline{SU-ITP-93-19}
\rightline{August 1993}
\rightline{hep-th/9308100}

\vfill

\title{Gedanken Experiments Involving Black Holes}

\vfill

\author{Leonard Susskind,\foot{susskind@dormouse.stanford.edu}
and L\'arus Thorlacius,\foot{larus@dormouse.stanford.edu}}

\vfill

\address{Department of Physics \break Stanford University, Stanford,
CA
94305-4060}

\vfill

\abstract
{\singlespace
Analysis of several gedanken experiments indicates that black hole
complementarity cannot be ruled out on the basis of known physical
principles.  Experiments designed by outside observers to disprove
the existence of a quantum-mechanical stretched horizon require
knowledge of Planck-scale effects for their analysis.  Observers who
fall through the event horizon after sampling the Hawking radiation
cannot discover duplicate information inside the black hole before
hitting the singularity.  Experiments by outside observers to detect
baryon number violation will yield significant effects well outside
the stretched horizon.}

PACS categories: 04.60.+n, 12.25.+e, 97.60.Lf

\vfill\endpage

\REF\hawking{S. W. Hawking
\journal Comm. Math. Phys. & 43 (75) 199
\journal Phys. Rev. & D14 (76) 2460.}

\REF\stu{L. Susskind, L. Thorlacius, and J. Uglum, {\it The Stretched
Horizon and Black Hole Complementarity,} Stanford University
preprint, SU-ITP-93-15, hepth/9306069, June 1993, to appear in Phys.
Rev. D.}

\REF\johnpri{J. Preskill, private communication at the ITP, UC Santa
Barbara Conference on ``Quantum Aspects of Black Holes'', June 21-26,
1993.}

\REF\thooftone{G. 't Hooft
\journal Nucl. Phys. & B335 (90) 138
\journal Physica Scripta & T36 (91) 247, and references therein.}

\REF\lenny{L. Susskind, {\it String Theory and the Principle of Black
Hole Complementarity,} Stanford University preprint, SU-ITP-93-18,
hepth/9307168, July 1993.}

\REF\unruh{W. G. Unruh
\journal Phys. Rev. & D14 (76) 870.}

\REF\thoofttwo{G. 't Hooft
\journal Phys. Rev. Lett. & 37 (76) 8
\journal Phys. Rev. & D14 (76) 3432
\journal Phys. Rev. & D18 (78) 2199.}

\REF\mempar{K. S. Thorne, R. H. Price, and D. A. MacDonald, {\it
Black
Holes:  The Membrane Paradigm},  Yale University Press, 1986, and
references therein.}

\REF\bek{J. D. Bekenstein
\journal Phys. Rev. & D9 (74) 3292.}

\REF\cbruhat{Y. Choquet-Bruhat
\journal Ann. Inst. Henri Poincar\'e & 8 (68) 327.}

\REF\page{D. N. Page, {\it Expected Entropy of a Subsystem,}
University of Alberta report, THY-22-93, gr-qc/9305007, May 1993.}

\chapter{Introduction}

One of the most interesting open questions in theoretical physics is
whether the unitary evolution of states in quantum theory is violated
by gravitational effects [\hawking].  The debate has centered on the
process of black hole formation and evaporation, where, at the
semi-classical level, almost no information about the initial quantum
state of the infalling matter appears to be carried by the final
outgoing Hawking radiation.  Our viewpoint in this paper is a
conservative one as far as quantum theory is concerned.  We shall
assume that outside observers can apply standard quantum mechanical
rules to the evolution of black holes.  This requires us to adopt a
more radical view of spacetime physics in the presence of a black
hole, but, as we shall argue, one which does not conflict with any
known laws of physics.

In reference [\stu] three basic postulates for the quantum evolution
of black holes were proposed.  The most important implication of
these postulates is that there exists a unitary $S$-matrix for any
such process.  Furthermore, the information carried by the infalling
matter re-emerges in the outgoing radiation and is not stored in a
stable or long-lived remnant.  A description by a distant observer,
necessarily involves non-trivial physical processes taking place in
the vicinity of the event horizon.  The nature of these processes is
such that organized energy and information is absorbed, thermalized
and eventually radiated.  Such behavior is commonly encountered in
macroscopic systems.  For example a cold piece of black coal which
absorbs a coherent laser signal will burn and emit the signal encoded
in thermal radiation.  The implication is that the ``stretched
horizon'' is described by an outside observer as a physical membrane
with microphysical structure and that the usual thermodynamics of
black holes follows from a coarse graining of the microscopic
description.

This point of view is in apparent contradiction with the expectation
that a freely falling observer encounters nothing unusual when
crossing the event horizon of a large black hole.  The duplication of
information behind the horizon and in the Hawking radiation seems to
violate the principles of quantum mechanics.  However obvious logical
contradictions only arise when one attempts to correlate the results
of experiments performed on both sides of the event horizon.  The
principle of black hole complementarity [\stu] states that such
contradictions never occur because the black hole interior is not in
the causal past of any observer who can measure the information
content of the Hawking radiation.

In this paper we shall illustrate the concept of black hole
complementarity by considering a number of gedanken experiments where
one might expect contradictions to arise.  The basic argument we
apply was formulated most clearly by J.~Preskill [\johnpri].  It says
that apparent contradictions can always be traced to unsubstantiated
assumptions about physics at or beyond the Planck scale.  We do not
offer a full ``resolution'' of the information paradox in this paper.
 Our aim is limited to challenging the commonly held view that, as
there is no strong curvature or other coordinate invariant
manifestation of the event horizon, an information paradox can be
posed without detailed knowledge of the underlying short distance
physics.  We analyze gedanken experiments which test the hypothesis
that the event horizon has no distinguishing feature to an observer
crossing it in free fall, while all information about the quantum
state of the infalling matter is returned to outside observers in the
Hawking radiation.  In each case we find that in order to expose a
violation of this hypothesis energies of order the Planck scale or
higher are required.  We conclude that the information paradox can
only be precisely formulated in the context of a complete theory of
quantum gravity and that the issue of information loss cannot be
definitively settled without such a theory.  We can, however, look
for clues to the nature of the final theory by making concrete
assumptions about information loss and exploring their consequences
in gedanken experiments.

The reference frame of an asymptotic observer and that of another
observer in free fall approaching the event horizon of a black hole
are very different.  As time measured by the distant observer goes
on, the boost relating the two frames becomes so enormous that an
electron at rest in one frame would appear to have super-planckian
energy in the other.  As t'~Hooft has stressed we have no
experimental experience of such energetic particles [\thooftone].
Their physical description may well be quite different from an
ordinary Lorentz boost applied to a localized object [\lenny].  We
wish to emphasize that black hole complementarity does not mean a
departure from the dictum that the laws of nature appear the same in
different frames of reference.  Rather, the assertion is that the
{\it description} of the same physical reality may differ quite
significantly between reference frames separated by a large boost
parameter.

We begin in Section~2 by comparing measurements made by different
observers in Rindler space.  In Section~3 we turn our attention to
black holes and discuss gedanken experiments in which outside
observers attempt to establish the non-existence of a physical
membrane at the stretched horizon.  We consider both static
geometries and a simple model of black holes formed by gravitational
collapse.  In Section~4 we analyze a more interesting class of
experiments involving observers who, after sampling the Hawking
radiation, cross the event horizon and attempt to observe duplicate
information.  These observers have to wait outside the black hole
until some of the information is returned and in Section~5 we give an
estimate of the time required.  In Section~6 we discuss the issue of
time-reversal in black hole evolution and Section~7 contains our
concluding remarks.

\chapter{Experiments in Rindler Space}

Before considering finite mass black holes it is instructive to
consider the simpler geometry of Rindler space which can be viewed as
the exterior region close to the horizon in the limit of infinite
black hole mass.  It is also isomorphic to the region of flat
four-dimensional Minkowski space defined by
$\vert z \vert  > \vert t  \vert$, $z>0$, shown in Figure~1.

We introduce Rindler coordinates $R$, $\omega$ as follows,
$$\eqalign{
t \; &= \; R \sinh{\omega} \;,\cr
z \; &= \; R \cosh{\omega} \;.\cr}
\eqn\rinco
$$
Note that $R$ is a spacelike coordinate and $\omega$ is time-like.
The Rindler line element is
$$
ds^2 \; = \;
-R^2\,d \omega^2 \; + \; dR^2 \; + \; dx^2 \; + \; dy^2 \;.
\eqn\rinmet
$$
The Rindler Hamiltonian generates translations of $\omega$, which are
Lorentz boosts of Minkowski space.  At time $t = 0$ it is given by
$$
H_R \; = \; \int_{-\infty}^\infty dx\,dy \int_0^\infty dz \;
z\; T^{00} (x, \; y, \; z)  \;,
\eqn\rinham
$$
where $T^{00}$ is the Minkowski space energy density.

In classical physics the evolution of fields in Rindler space can be
described in a self contained fashion without any reference being
made to the regions of Minkowski space that have been excised in
Figure~1.  Events which take place in the region $z-t>0$, $z+t<0$ can
be summarized by initial conditions at $\omega =-\infty$, and the
other excised regions are not in the causal past of the Rindler
space.

\vbox{
\vskip 20pt
{\centerline{\epsfsize=3.0in \epsfbox{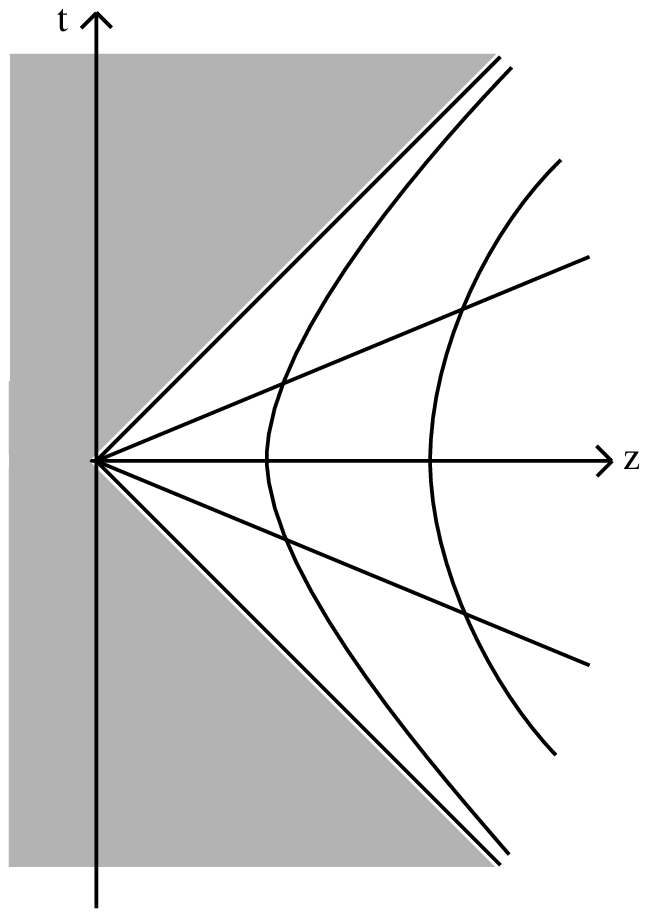}}}
\vskip 12pt
{\centerline{\tenrm FIGURE 1.}}
\vskip 12pt
{\centerline{\vbox{\hsize 5in \singlespace\tenrm \noindent
Rindler space as a wedge of Minkowski space: surfaces of constant
Rindler time are straight lines through the origin while surfaces of
fixed Rindler position are hyperbolas.  There is a future (past)
event horizon at $z{-}t{=}0$ ($z{+}t{=}0$).
}}}
\vskip 15pt}

When dealing with quantum fields, the space of states for a Rindler
observer consists of the functionals of fields defined on the
positive $R$ axis.  This restriction requires the usual Minkowski
space vacuum to be described in Rindler space as a density matrix for
a mixed state.  The appropriate density matrix was obtained by Unruh
[\unruh] and is given by
$$
\rho \; = \; \exp{(-2 \pi H_R)} \;.
\eqn\dense
$$
The form of the Rindler density matrix corresponds to a thermal state
with a temperature $T_R = {1 \over 2\pi}$.  Although, \dense\ is
merely a rewriting of the Minkowski vacuum, the thermal effects are
nevertheless quite real to an accelerated observer.  For example any
standard thermometer at rest in Rindler coordinates at position $R$
will record a temperature ${1 \over 2 \pi R}$\foot{The factor of ${1
\over R}$ in the temperature comes from the conversion between
Rindler coordinate time and the proper time registered by a clock
attached to the thermometer.}.  The Minkowski observer attributes the
same effect to the thermometer's uniform acceleration of magnitude
${1 \over R}.$  More generally, observers at rest in Rindler
coordinates will describe the Minkowski vacuum as a state with
position-dependent proper temperature.  For large values of $R$ the
acceleration of such an observer is small and the proper temperature
goes to zero asymptotically, while Rindler observers near the edge of
the Rindler wedge experience enormous acceleration and measure a
temperature which diverges as $\vert z \vert \rightarrow \vert t
\vert.$

It is interesting to consider temperature dependent phenomena such as
baryon number violation at different locations in Rindler space.  At
first sight one might expect significant B-violation due to the `t
Hooft anomaly [\thoofttwo] when the local proper temperature reaches
the weak scale.  However, this is not so.  The temperature only
reaches or exceeds a given value, $T$, over a region of proper size
$T^{-1}$ in the $R$ direction.  The gauge field configurations which
ordinarily lead to B-violation at high temperature have a much larger
size of order ${1 \over g^2 T}$ and therefore cannot
contribute.\foot{The infinite extension in the x, y directions does
not affect this.}  Configurations of scale ${1 \over T}$ contribute
with the usual exponential supression, $\exp{(-{4 \pi \over g^2})}$.
On the other hand, when the local temperature reaches the GUT scale,
$T_{GUT}$, $X$ bosons of wavelength $T_{GUT}^{-1}$ will mediate
B-violation, suppressed only by powers of the coupling $g$.

Suppose that a Rindler observer wishes to test the prediction that
baryon number violation takes place within a distance of order
$T_{GUT}^{-1}$ from the edge of Rindler space.  The observer prepares
a sealed ``bucket'' with walls that can transmit heat but not baryon
number\foot{Although the manufacture of such buckets from
conventional matter may be impossible, their existence does not
violate the rules of relativistic quantum field theory.}.  The baryon
number in the bucket is measured and recorded at the beginning of the
experiment.  The bucket is then lowered toward $R=0$ and retrieved
after its bottom edge reaches $R=T_{GUT}^{-1}$, as indicated in
Figure~2.  The baryon number is remeasured and compared with its
original value.  According to the Rindler observer the thermal
effects can induce a change in $B$.  The same phenomenon is described
by a Minkowski observer as due to the intense acceleration disturbing
the virtual processes involving X-bosons inside the bucket.

By contrast, a Minkowski observer would see no baryon number
violation for a freely falling bucket which falls through the Rindler
horizon at $z-t=0$.  This does not contradict the result of the
previous experiment for several reasons.  First of all, the absence
of B-violation cannot be reported back to the Rindler observer.  One
might try to communicate the absence of B-violation at the bottom of
the freely falling bucket during the time that the bottom is between
$R=T_{GUT}^{-1}$ and $R=0$.  To do so will require the message sent
by the bucket to be composed of quanta of frequency at least
$T_{GUT}$ as seen in the frame of the bucket.  Therefore the
processes, by which the message is sent, will themselves lead to
baryon number violation.  If lower frequencies are used the
uncertainty principle will insure that the message will be lost
behind the Rindler horizon.

\vbox{
\vskip 20pt
{\centerline{\epsfsize=3.0in \epsfbox{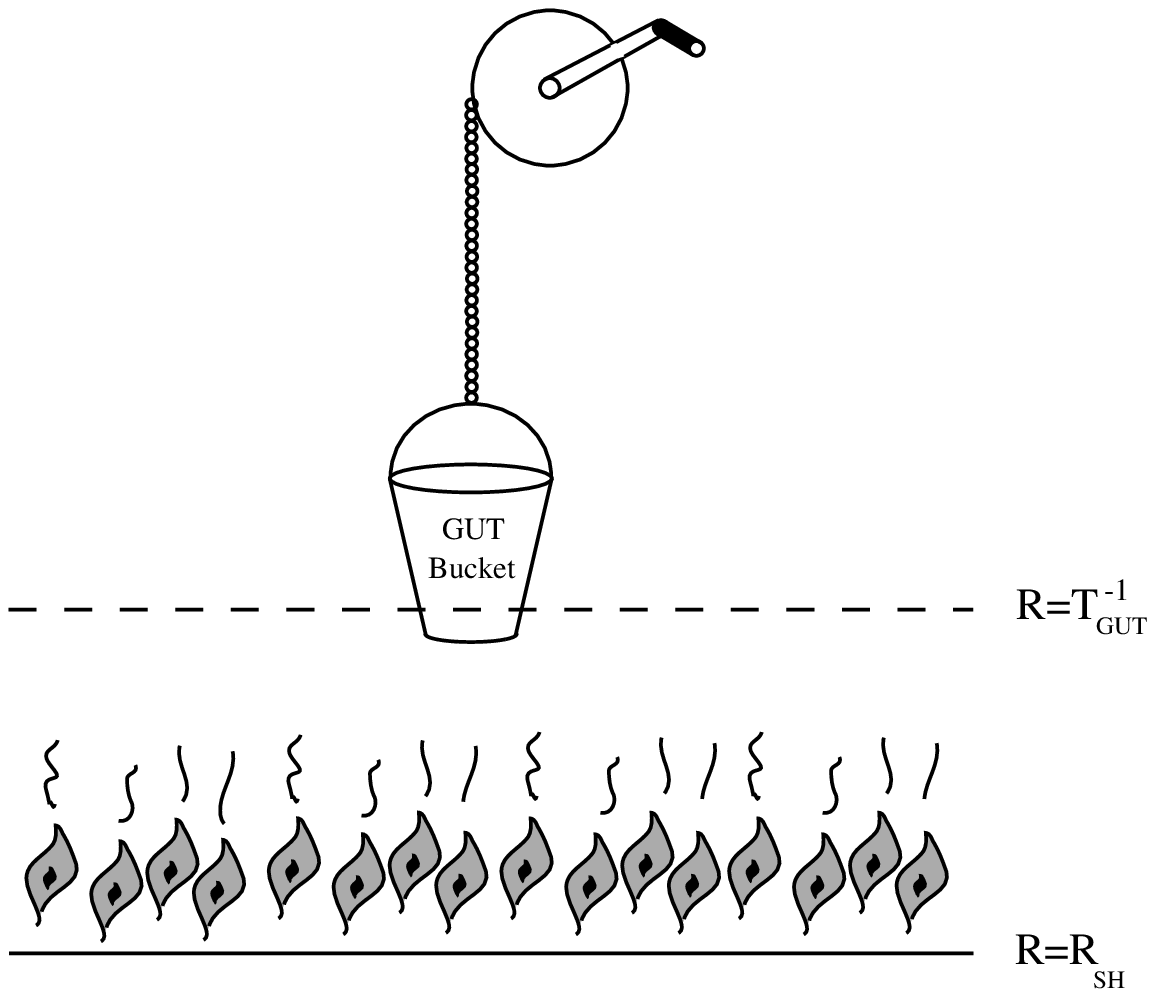}}}
\vskip 12pt
{\centerline{\tenrm FIGURE 2.}}
\vskip 12pt
{\centerline{\vbox{\hsize 5in \singlespace\tenrm \noindent
A gedanken experiment to test baryon nonconservation near the Rindler
horizon.
}}}
\vskip 15pt}

Furthermore there is a real sense in which baryon number violation
does take place in the frame of the Rindler observer even for the
freely falling bucket.  B-violation is of course continuously taking
place in short lived processes in the freely falling frame of the
bucket.  However everyday observations average over much longer time
scales and do not resolve these virtual effects.  Consider now
averaging the baryon number over a finite resolution of the Rindler
time $\omega$.  As the event horizon is approached a finite time
interval in the freely falling frame becomes indefinitely stretched
in Rindler time.  A virtual fluctuation,  where an X-boson is emitted
causing a transition from a down quark to an electron on one side of
the horizon and reabsorbed on the other side, appears to the Rindler
observer as a violation of baryon number.  The recombination is never
observed in Rindler space.

Even closer to $R=0$ we encounter temperatures approaching the Planck
scale where the laws of nature are unknown.  Therefore we can only
deal with this region phenomonogically.  From the point of view of a
distant Rindler observer it can be replaced by a membrane or
stretched horizon with certain properties such as electrical
conductance and vanishing reflectivity [\mempar].  Most importantly
it has a proper temperature of planckian magnitude.  It provides the
hot boundary which the rest of the Rindler space is in equilibrium
with.

Let us consider matter in a particular quantum state which falls
through the Rindler horizon.  Rindler observers can describe this
process in the following way.  The infalling matter approaches the
stretched horizon and is absorbed.  At this point the Rindler
observer may conjecture the existence of Planck scale degrees of
freedom on the stretched horizon which become correlated with the
initial quantum state of the absorbed matter and store the
information among a large number of thermalized degrees of freedom.
The quantity of information that can be indefinitely stored is
infinite because the area of the stretched Rindler horizon is
infinite.  To a distant observer this is indistinguishable from
information loss.

We now turn to some simple gedanken experiments that the Rindler
observer can do to try to discover whether the infalling information
is actually stored at the stretched horizon.  Our first example
involves an observer who approaches and examines the stretched
horizon and then returns to report the results (see Figure~3).  Any
observer who penetrates all the way to the stretched horizon will
have to undergo Planck scale acceleration to return.  As a result
this experiment cannot be analyzed in terms of known physics and
therefore it cannot at present be used to rule out the Rindler
observer's hypothesis that information is stored at the stretched
horizon.

Next consider an experiment in which a freely falling observer, who
passes through the event horizon, attempts to continuously send
messages to the outside reporting the lack of substance of the
membrane.  First suppose his messages are carried by radiation of
bounded frequency in the free-falling frame.  Because the observer
has only a finite proper time before crossing the Rindler event
horizon only a finite number of bits of information can be sent.  The
last few bits get enormously stretched by the red shift factor and
are drowned by the thermal noise.  Therefore there is in a sense a
last useful bit.  If the carrier frequency is less than the Planck
frequency the last useful bit will be emitted before the stretched
horizon is reached.  In order to get a message from behind the
stretched horizon the observer must use super-planckian frequencies.
Again the experiment cannot be analyzed using conventional physics.

\vbox{
\vskip 20pt
{\centerline{\epsfsize=3.0in \epsfbox{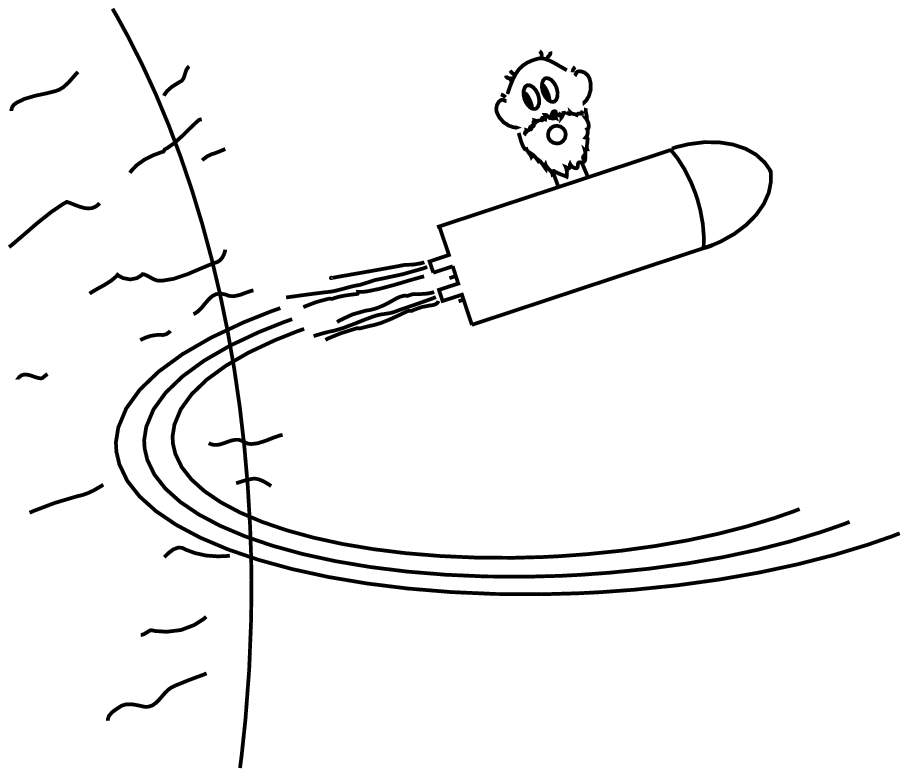}}}
\vskip 12pt
{\centerline{\tenrm FIGURE 3.}}
\vskip 12pt
{\centerline{\vbox{\hsize 5in \singlespace\tenrm \noindent
To approach the stretched horizon and return, an observer must
undergo Planck scale acceleration.
}}}
\vskip 15pt}

In both these experiments, efforts made to investigate the physical
nature of the stretched horizon are frustrated by our lack of
knowledge of Planck scale physics.  J.~Preskill has speculated that
this is a general feature of all such experiments and as a result
there may be no well posed information paradox in black hole physics
[\johnpri].

\chapter{Experiments Outside Black Holes}

The event horizon of Rindler space can be viewed as a black hole
event horizon in the limit of infinite mass. Any attempt by external
observers to determine the physical presence of an information
storing membrane outside a black hole event horizon will run into
planckian obstacles just as in Rindler space.  The discussion of the
gedanken experiments of the previous section carries over to this
case as the reader can easily check.  In this case we define the
stretched horizon as the time-like surface where the area of the
transverse two-sphere is larger than at the null event horizon by
order one in Planck units.  With this definition the proper
acceleration of a point on the stretched horizon at fixed angular
position is approximately one Planck unit.  For a static Schwarzchild
geometry the proper spatial distance from the stretched horizon to
the event horizon is also about one Planck unit.

Two important differences do occur when one considers black holes of
finite mass.  The first is that the red shift factor between the
stretched horizon and infinity is finite and an asymptotic observer
sees finite temperature radiation.  This leads to the evaporation of
the black hole.  For large black holes the evaporation is very slow
and one can approximate the evolving geometry by a static one for the
purpose of discussing messages sent by observers, who come close to
or enter the stretched horizon.  The second difference, granting that
the stretched horizon can store quantum information, is that the
storage capacity is finite and given by the Bekenstein entropy
[\bek].  This implies that the quantum correlations of the initial
state are returned in the Hawking radiation [\stu].  Furthermore, a
stretched horizon which carries finite entropy cannot hold onto the
quantum information for an indefinite length of time.  We shall
discuss further the concept of a finite information retention time in
the following sections.

Now consider a non-static geometry corresponding to the formation of
a large black hole by infalling matter.  The simplest case to study
involves a thin spherical shell of massless matter [\unruh,\cbruhat].
 The geometry is constructed by matching a flat spacetime and a
massive Schwarzchild solution across a radial null surface as in
Figure~4.

In order to discuss gedanken experiments performed by outside
observers it is convenient to define coordinates which cover the
exterior of the black hole.  Near $\scripm$ the geometry approaches
flat spacetime.  We introduce polar coordinates in this region
denoted by $(t,r^*,\theta,\varphi)$.\break
\vbox{
\vskip 20pt
{\centerline{\epsfsize=3.0in \epsfbox{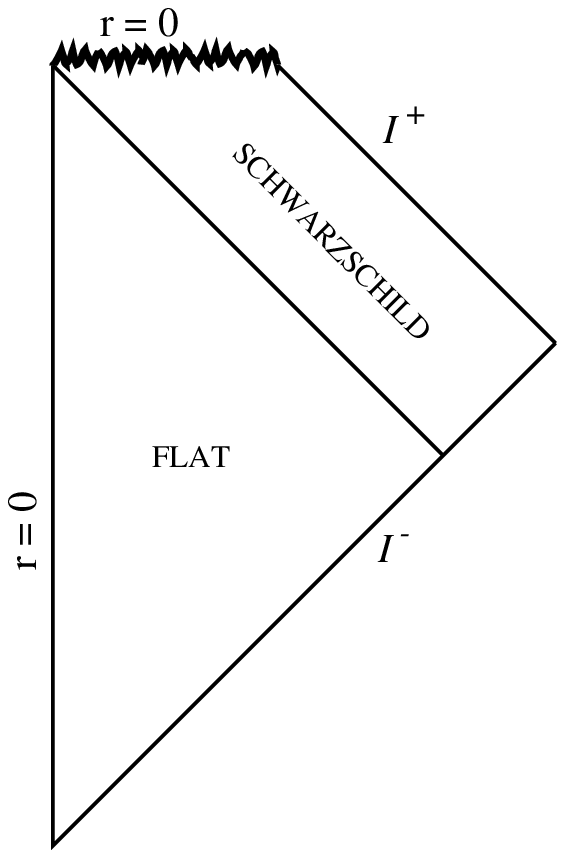}}}
\vskip 12pt
{\centerline{\tenrm FIGURE 4.}}
\vskip 12pt
{\centerline{\vbox{\hsize 5in \singlespace\tenrm \noindent
Penrose diagram for a geometry describing an infalling shell of
light-like matter.
}}}
\vskip 15pt}
The combinations $u=t-r^*, v=t+r^*$ are null coordinates, and can be
continued by extrapolating the light cones to finite locations.
$r^*$ and $t$ can then be defined in the entire region outside the
event horizon as
$$\eqalign{
t \; &= \; {v+u \over 2}\;,\cr
r^* \; &= \; {v-u \over 2} \;.\cr}
\eqn\tort
$$
The global event horizon shown in Figure~5 is the null surface
$u=\infty$ which determines the boundary of the black hole region.

As in the static case a stretched horizon can be defined as a time
like surface just outside the event horizon.  This can be achieved in
a number of ways.  The method proposed in [\stu] is as follows.  At a
point on the global event horizon construct the radial null ray which
does not lie in the horizon.  That ray intersects the stretched
horizon at a point where the area of the transverse two-sphere has
increased by an amount of order a Planck unit relative to its value
at the corresponding point on the event horizon.  An important new
feature of the collapse geometry as compared to an eternal black hole
is that both the event horizon and the stretched horizon extend into
the flat spacetime region inside the infalling matter shell.  The
stretched horizon begins (see Figure~5) at the same finite value of
$v$ at which the event horizon initially forms.  At this point the
area of the stretched horizon is approximately one Planck unit and it
is not useful to extend its definition beyond that.

\vbox{
\vskip 20pt
{\centerline{\epsfsize=3.0in \epsfbox{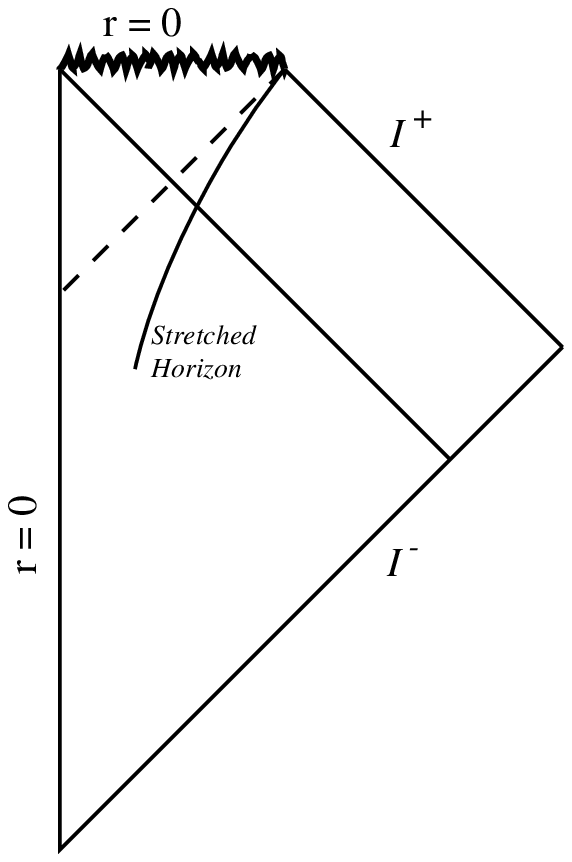}}}
\vskip 12pt
{\centerline{\tenrm FIGURE 5.}}
\vskip 12pt
{\centerline{\vbox{\hsize 5in \singlespace\tenrm \noindent
Location of horizons in a collapse geometry.  The dashed line
represents the global event horizon.
}}}
\vskip 15pt}

An important property of the stretched horizon is that the proper
acceleration is about unity in Planck units everywhere on it,
independent of time.  This means that the stretched horizon can again
be viewed as a hot boundary with planckian temperature.  The red
shift factor between the portion of the stretched horizon below $v=0$
and infinity can be computed.  The result is that the quanta reaching
infinity have energy of order ${1 \over M}$ where $M$ is the total
infalling mass.  This agrees with other calculations.  One might
object that an observer in the flat spacetime region, who is unaware
of the incoming shock wave, should not detect radiation coming from
the stretched horizon.  This, however, depends on the state of motion
of the observing apparatus.  In particular, if the apparatus is at
rest with respect to the tortoise coordinates \tort\ then it is
subject to a proper acceleration and will register radiation.

We are interested in testing the hypothesis that, from the outside
viewpoint, infalling quantum information is stored on the stretched
horizon and subsequently emitted during the evaporation process.
More precisely, we would like to show that any such test runs up
against our ignorance of Planck scale physics.  Gedanken experiments
involving messages sent by probes which enter the black hole will be
frustrated by the Planck scale red shift factor from the stretched
horizon to infinity, in precisely the same way as in the case of a
static geometry.  If the signal frequency is below the Planck
frequency the last useful bit of information will be emitted before
the probe reaches the stretched horizon.  This is true even for a
probe which is sent into the black hole in the flat region of
spacetime inside the imploding shell.  Similarly, probes which
approach the stretched horizon, and then escape to infinity, will
necessarily experience planckian acceleration at some point with
unforeseeable consequences.

\chapter{Experiments Performed Inside Black Holes}

In this section we shall consider a class of experiments which
explore quantum correlations between events on both sides of the
event horizon.  In particular, we are interested in contradictions
which arise from the apparent duplication of quantum information.  If
the Hawking radiation is to faithfully encode the infalling
information, while the infalling matter crosses the event horizon
without disruption, one might be led to infer such duplication
[\stu].  This would violate the linearity of quantum mechanics.  So
long as any single observer can never learn the results of
experiments performed on both copies of the quantum state we are not
led to logical contradiction.  It is clear that an observer, who
never enters the black hole, can know nothing of experiments
performed inside.  We must, however, also consider observers who
enter the black hole and insist that they do not experience any
violation of quantum mechanics.  Indeed, for all we know, we
ourselves could be living inside the event horizon of a gigantic
black hole at this very moment.

Let us first consider a simple experiment, indicated in Figure~6,
involving measurements made on both sides of the event horizon of a
large black hole of mass $M$.  A pair of ``spins''\foot{By spin we
mean an internal label not coupled to a long range gauge or
gravitational field.} are prepared in a singlet state at point $A$
and then one of them is carried into the black hole.  Two spin
measurements are made at points $B$ inside the event horizon and $C$
outside.  The results are then transmitted to an observer at $D$ who
can establish the correlation between them.\foot{The single spin pair
could be replaced by an ensemble of identically prepared pairs in
order to measure statistical quantum correlations.}  No Planck scale
physics enters into the analysis of this experiment but neither does
it lead to any paradox since no observation was made on the Hawking
radiation.

\vbox{
\vskip 20pt
{\centerline{\epsfsize=3.0in \epsfbox{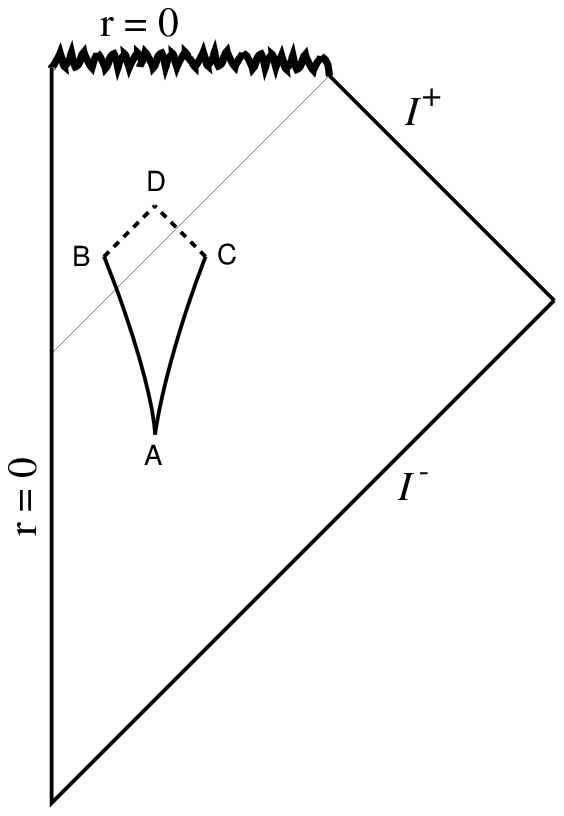}}}
\vskip 12pt
{\centerline{\tenrm FIGURE 6.}}
\vskip 12pt
{\centerline{\vbox{\hsize 5in \singlespace\tenrm \noindent
A gedanken experiment involving correlations on both sides of the
event horizon.  A pair of correlated ``spins'' are created at $A$.
The spins are measured at $B$ and $C$ and the results are
communicated to an observer at $D$.
}}}
\vskip 15pt}

We turn now to a class of experiments distinguished by the existence
of an observer who first performs measurements on outgoing Hawking
radiation and then falls through the event horizon.  Typically the
observer then receives a signal from some system which previously
fell through.  In this way the observer may collect duplicate
information which could potentially lead to contradictions.

As an illustrative example, indicated in Figure~7, consider an
experiment in which a pair of particles is prepared in a ``spin''
singlet.  One member $a$ of the pair is sent into a black hole along
with an apparatus $A$ which can measure the spin and send out
signals.  The other member $b$ remains outside.  For definiteness we
assume that the energy associated with the apparatus is small
compared to the black hole mass $M$ and that it is initially at rest
outside the black hole.

\vbox{
\vskip 20pt
{\centerline{\epsfsize=3.0in \epsfbox{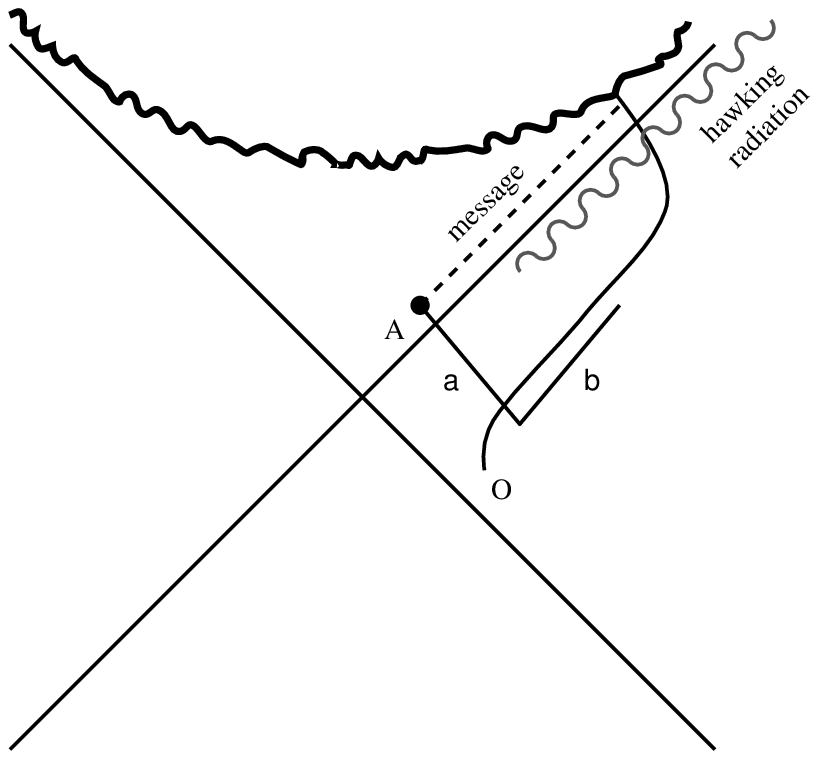}}}
\vskip 12pt
{\centerline{\tenrm FIGURE 7.}}
\vskip 12pt
{\centerline{\vbox{\hsize 5in \singlespace\tenrm \noindent
A gedanken experiment in which an observer $O$ measures information
in Hawking radiation before falling into the black hole.  A spin $a$
has previously crossed the horizon and is measured by apparatus $A$.
A message is sent from $A$ to $O$ before $O$ hits the singularity.
}}}
\vskip 15pt}

An observer $O$ who has been hovering outside the hole makes
measurements on the Hawking radiation.  Assuming that all infalling
information is eventually radiated, a measurement can be performed on
the radiation which is equivalent to a determination of any component
of the original spin.  Meanwhile the infalling spin $a$ has been
measured by the apparatus $A$ which accompanied it.  From the point
of view of an external observer the ``spin in the Hawking radiation''
$h$ must be correlated with the member $b$ of the original pair which
remained outside the black hole.  If the spin $b$ is measured along
any axis, then the Hawking spin $h$ must be found anti-aligned if it
too is measured along the same axis.  On the other hand the original
spin $a$ which fell through the horizon was also correlated to the
other member of the pair $b$.  It would seem that the two separate
spins ($a$ and $h$) are correlated with a third ($b$) so as to be
anti-aligned with it.  This violates the principles of quantum
mechanics.

In reference [\stu] it was argued that no logical contradiction could
be derived from this circumstance because no observer could know the
results of the measurements of both the spin behind the horizon and
the Hawking spin.  An observer who measures the Hawking spin can,
however, subsequently fall through the horizon carrying a record of
his measurements, as was pointed out to us by J. Preskill [\johnpri].
 If a message can be sent to that observer by the apparatus which
measures the infalling spin then he clearly discovers a violation of
quantum mechanics.  The resolution of this difficulty was suggested
by Preskill himself.  We shall see that if the observer $O$ falls
through after enough time has elapsed for the relevant quantum
information to be gathered and if he is to receive the message from
$A$ before hitting the singularity, the message must be sent in
quanta of energy far beyond the Planck scale.

In the following section we shall discuss the length of time a piece
of information remains on the stretched horizon before being emitted.
 We shall give plausible arguments that this information retention
time is typically of order $M^3$.  For now, we assume that a distant
observer will have to conduct experiments on the outgoing Hawking
radiation for a period $\tau \sim M^3$ before information can be
recovered.  During this time the mass of an evaporating black hole
significantly decreases (the total lifetime is also of order $M^3$)
but we shall ignore this for the time being and make use of a static
Schwarzschild geometry.  It will become apparent that the effect of
evaporation is to strengthen our conclusion.

Since we wish to discuss observations made inside a black hole it is
convenient to use Kruskal coordinates which extend past the event
horizon,

$$\eqalign{
U \; &= \; -\exp{
\bigl({r^* - t \over 4M}\bigr)} \;,\cr
V \; &= \; \phantom{-}\exp{
\bigl({r^* + t \over 4M}\bigr)} \;.\cr}
\eqn\kruskal
$$
The Schwarzschild line element becomes
$$
ds^2 \; = \; - {32M^3\exp{({-r \over 2M})}\over r} \; dUdV \;.
\eqn\metric
$$
The future event horizon is at $U=0$ and the singularity is at
$UV=1$.  The geometry is shown in Figure~7.  It is evident that the
value of $U$ where the observer $O$ runs into the singularity becomes
very small if the observer delays for a long time entering the black
hole.  This in turn constrains the time which the apparatus $A$ has
available to emit its message.  Let us choose the origin of our
tortoise time coordinate so that the apparatus passes through the
stretched horizon at $V=1$.  The observer $O$ will go through the
stretched horizon after a period of order $M^3$ has passed in
tortoise time, i. e. at  $\log{V} \sim M^2$.  Since the singularity
is at $UV=1$ the message from $A$ must be sent before the apparatus
reaches $U \sim \exp{(-M^2)}$.  Near $V=1$ this corresponds to a very
short proper time $\tau \sim M^2 \exp{(-M^2)}$.  The uncertainty
principle then dictates that the message must be encoded into
radiation with super-planckian frequency $\omega \sim
M^{-2}\exp{(M^2)}$.  The back-reaction on the geometry due to such a
high-energy pulse would be quite violent.  It is apparent that the
apparatus $A$ cannot physically communicate the result of its
measurement to the observer $O$ in this experiment.

One can imagine a number of gedanken experiments which are variations
on this theme.  They all lead to the conclusion that attempts to
evade black hole complementarity involve unjustified extrapolation
far beyond the Planck scale.  The evaporation of the black hole
modifies the geometry in a manner which only makes the time available
for $O$ to receive the message shorter.

\chapter{Information Retention Time}

We wish to define a measure of the ``lifetime'' of information stored
in a thermalized system, $K$, such as a black hole stretched horizon
or a thermal cavity.  Suppose at time $t=0$, the system $K$ is in
some pure state $\psi$ which is a typical member of a thermal
ensemble with energy $E$ and temperature $T$.  Furthermore, let
$N(E)$ be the number of possible states at energy $E$.  To begin
with, the system has zero fine grained entropy and coarse grained
entropy equal to $\log{N}$.

We assume that the system $K$ radiates thermal radiation for a time
$t$.  At the end of this time $K$ is no longer in a pure state since
it is correlated to the outgoing radiation field.  Both the radiation
and the system are described by density matrices and from these an
entropy of entanglement can be defined [\stu].  This entropy is the
same for both subsystems and is given by
$$
S_E \; = \; - {\rm Tr}\, \rho_K \log{\rho_K} \;.
\eqn\entrpy
$$
Eventually this entropy will return to zero when all the heat has
been radiated and the system $K$ has reached its ground state.

The information contained in the radiation at time $t$ is defined to
be
$$
i(t) \; = \; S_{max} - S_E (t) \;,
\eqn\info
$$
where $S_{max}$ is the thermal entropy of the radiation calculated as
if it were emitted as conventional black body radiation.  It is
roughly equal to the number of emitted photons.  D.~Page's
calculation of the entropy of a subsystem [\page] implies that the
information $i(t)$ remains extremely close to zero until the number
of emitted photons is of order ${1 \over 2} \log{N}$.  In other words
no information is radiated until the thermal entropy of $K$ has
decreased to half its original value.  After this time the
information in the radiation increases linearly with the number of
emitted photons.  Thus there is a considerable length of time before
any information is released.  In the case of a black hole of initial
mass $M$ it is equal to the time it takes the horizon area to
decrease to half its initial value.  A simple calculation shows this
{\it information retention time} to be $\tau_i \sim M^3$.

One may also consider the information retention time for a black hole
which is being prevented from evaporating by an incoming energy flux.
 It seems likely that this would increase the information retention
time because it would increase the number of available degrees of
freedom among which the information is shared.

An information time of order $M^3$ leaves a comfortable margin in the
analysis of the gedanken experiment in the previous section.  Turning
the argument around, one can use black hole complementarity to obtain
a lower bound on the information retention time.  As long as $\tau_i
\gsim M \log{M}$ the observing apparatus $A$ will have to use at
least Planck scale frequency to get a message to the observer $O$,
who enters the black hole after making measurements on the Hawking
radiation.  If $\tau_i$ were to be any shorter, then one could detect
a violation of quantum mechanics in such an experiment for a
sufficiently massive black hole.

\chapter{Experiments Involving Time Reversal}

The time reversal of the formation and evaporation of a black hole
has been a subject of controversy.  Confusion about white holes and
CPT easily arises when one contemplates information loss.  This
confusion is straightforwardly resolved in a theory with a quantum
mechanical stretched horizon.  Let us consider a black hole formed
from the collapse of a diffuse cloud of elephants.  The evaporation
products consist of an outgoing train of thermal radiation whose
temperature increases with time from ${1 \over 8 \pi M}$ initially to
temperatures approaching the Planck scale at the end.  The postulate
of the validity of quantum mechanics  predicts that if this radiation
is time-reversed it will evolve back into an expanding diffuse cloud
of elephants.  We can consider such an experiment to be a test of the
postulates of reference [\stu].

Compare this with the result that would be obtained in the
semi-classical approximation in which the time reversed average
energy flux is used as a source of gravitation.  In this case any
initially formed small black hole will itself Hawking radiate and
prevent the buildup to a large black hole.  Furthermore, even if a
large black hole were to form, there would be severe Boltzman
supression of finding even a single elephant in the final state.
Thus a semiclassical analysis inevitably leads to information loss.

According to the view expressed in reference [\stu] the experiment
described above is not fundamentally different from the explosion of
a bomb among a herd of elephants.  Again there is no chance of
reconstructing the initial state from any crude approximation to the
time reversed final state or its evolution.  The conventional
understanding of this type of experiment is that only a very tiny
subset of configurations will evolve back to the kind of organized
state that we started with.  Very minute disturbances will cause the
time-reversed state to evolve instead into a typical disorganized
thermal state.

As with the other gedanken experiments described here, detailed
analysis of the above black hole experiment cannot be carried out
without the knowledge of Planck scale physics.  This is evident from
the fact that the first thing to happen in the time reversed process
is the collision of Planck energy particles to form a small black
hole. Our inability to follow even this early evolution renders any
discussion of the remainder futile.  While the assumption of a
quantum-mechanical stretched horizon does not in itself provide a
detailed dynamical evolution of the time reversed process it is
unambiguous in predicting that a sufficiently accurate construction
of the time reversed state will lead to elephants emerging from a
white hole.

The following extreme example illustrates how misleading
semiclassical considerations can be.  Consider a very high energy
particle in an incoming $S$-wave in otherwise empty space.  The wave
function has support on a thin infalling shell.  According to the
usual semiclassical rules we first compute the expectation value of
the energy momentum tensor.  In this case it would describe a thin
incoming spherical shell.  We would then use that as a source in
Einstein's equations and solve for the combined evolution of the
shell and the geometry.  If the initial energy is well above the
Planck scale a black hole will form in this approximation.  The next
step in a semi-classical calculation is to account for quantum
effects, such as Hawking radiation, in the background geometry.  When
back-reaction is included, the semiclassical method further predicts
that the black hole evaporates into thermal radiation.  This is
almost certainly not what happens.  A single particle does not
gravitationally attract itself and cannot form a black hole.  This
means that further corrections to the semi-classical approximation
must lead to an entirely different picture of the final state.  Once
again our inexperience with Planck scale physics precludes detailed
systematic improvement of the semi-classical picture.  However in
this rather trivial example it seems almost certain that the incoming
$S$-wave in fact evolves into an outgoing $S$-wave state of a single
particle.

\chapter{Summary}

In this paper we have analysed a number of gedanken experiments to
illustrate the consistency of the following two hypotheses:

1)  To an outside observer all information falling onto a black hole
is stored and thermalized on the stretched horizon until it is
radiated during evaporation.  The thermalization involves processes
which wash out baryon number.

2)  To an infalling observer no exceptional phenomenon such as
information bleaching or significant baryon number violation will
take place upon crossing the horizon.

An observer hovering near the horizon experiences enormous proper
acceleration and sees intense radiation emanating from the black
hole.  This observer can for most purposes replace the black hole by
a hot membrane located just outside the event horizon.  Our examples
suggest that external experiments designed to detect whether quantum
information is stored at the stretched horizon cannot be analyzed
using presently known physics.  Furthermore experiments performed
behind the horizon but far from the singularity cannot detect
information duplication.  This leads us to question whether Hawking's
information paradox can be well posed at present.

In addition to adressing the information question the stretched
horizon concept provides insight into other puzzles.  It demands
that, to outside observers, baryon number violation appears to take
place at a distance of order $M_{GUT}^{-1}$ from the event horizon
and is not hidden behind the horizon.

The issue of time reversability of black hole processes is confusing
in the context of information loss.  By contrast, if it is admitted
that a quantum-mechanical stretched horizon exists, standard
statistical reasoning applies to black holes.  Time reversed black
holes or white holes are possible but no more likely than the random
motion in a swimming pool ejecting a diver up to a spring board.

It is our view that black hole complementarity is not derivable from
a conventional local quantum field theory.  It seems more likely that
it requires a radically different kinematical description of physics
at very high energy, such as string theory [\lenny].  Our point in
this paper is that the revolutionary new elements need only become
manifest at extreme energies where our present knowledge is
insufficient.  We may, however, be able to use black hole
complementarity to guess some features of the new kinematics.

\underbar{Acknowledgements:} \hfill\break\indent
Many of the ideas presented here grew out of discussions we had with
various participants in the conference on Quantum Aspects of Black
Holes at the ITP, UC Santa Barbara, June 21-26, 1993.  In particular
we would like to thank T.~Jacobson, D.~Page, A.~Peet, G.~t'~Hooft,
J.~Uglum, H.~Verlinde, R.~Wald and especially J.~Preskill for useful
conversations.  We thank J.~Susskind for assistance in preparing the
manuscript.  This work was supported in part by National Science
Foundation grant PHY89-17438.

\refout
\end